\documentstyle[prd,aps,preprint,12pt]{revtex}
\tightenlines
\input{epsf.sty}
\input{psfig.sty}

\newcommand{\be}{\begin{equation}}
\newcommand{\ee}{\end{equation}}
\newcommand{\bea}{\begin{eqnarray}}
\newcommand{\beas}{\begin{eqnarray*}}
\newcommand{\eea}{\end{eqnarray}}
\newcommand{\eeas}{\end{eqnarray*}} 
\newcommand{\ba}{\begin{array}}
\newcommand{\ea}{\end{array}}
\begin{document}

\draft
\preprint{\vbox{\hbox{IFT-P-029/2001 }}}

\title{Small Active and Sterile Neutrino Masses from the TeV Scale} 

\author{
A. P\'erez-Lorenzana$^{1,2}$\footnote{e-mail:aplorenz@ictp.trieste.it} 
and C. A. de S. Pires$^{3}$\footnote{e-mail:cpires@ift.unesp.br}}
\address{
$^1$ The Abdus Salam International Centre for Theoretical Physics, I-34100,
Trieste, Italy\\
$^2$  Departamento de F\'\i sica,
Centro de Investigaci\'on y de Estudios Avanzados del I.P.N.\\
Apdo. Post. 14-740, 07000, M\'exico, D.F., M\'exico\\ 
$^3$ Instituto de F\'\i sica Te\'orica,
Universidade Estadual Paulista.\\
Rua Pamplona 145, 01405-900- S\~ao Paulo, SP
Brazil.}

\date{August, 2001}

\maketitle
\begin{abstract}
A new mechanism for understanding small neutrino masses using
only simple new physics at the TeV scale is proposed. 
As an application, it is shown how it can naturally lead 
to the mass hierarchy of the so called bimaximal mixing in the case of three
active neutrinos, or  the  (3+1) scenarios for sterile neutrinos, 
using only the $SU(2)_L$ quantum numbers of the particles.\\[1em]
PACS:14.60.Pq; 14.60.St.
\end{abstract}

\section{Introduction}

The most commonly used way to understand the smallness of neutrino
mass~\cite{review}
compared to that of the charged fermions is the seesaw
mechanism\cite{seesaw}, according to which one has,
\begin{eqnarray}
{\cal M}_{\nu} \simeq -M^T_DM^{-1}_RM_D .
\end{eqnarray}
For choices of the Dirac mass matrix $M_D$ dictated by simple unified
models, one expects the typical scale of the seesaw masses to be around
$10^{10}$ GeV or higher for neutrino masses required to fit
atmospheric and solar data~\cite{exp1,exp2}. 
In this class of models, neutrino masses and 
mixings are dictated by physics at superheavy scales. An important
question to ask (and has often been discussed in literature) therefore is
whether the neutrino masses can arise from new physics 
nearby the weak scale, e.g., in the TeV range so that one may hope to probe
them in the collider or other indirect experiments.

There exist several TeV scale scenarios for neutrino masses in the
literature: (i) radiative scenarios that use extensions of the standard
model by extra charged Higgs bosons\cite{babu} (ii) scenarios using
bulk neutrinos~\cite{dienes,mnp,others} or bulk scalars~\cite{ma1,carlos} 
and (iii) a more recent one \cite{ma2} that
uses the seesaw mechanism but with naturally  suppressed Dirac
masses $M_D$.

In this letter, we propose a new mechanism that uses higher
dimensional operators~\cite{weinberg} that lead to a naturally suppressed
neutrino masses using only simple new physics at the TeV scale. The higher
dimensional operator could arise in a number of ways from TeV scale
physics depending on the nature of extra symmetries and the
particle content of the theory. We will illustrate our mechanism using
the effective low energy theory at the weak scale.

For simplicity, 
we will assume the low energy theory to be the standard model plus an
extra Higgs doublet and a Higgs singlet field. We will assume also the theory
to have softly broken global symmetries such as $L=L_e+L_\mu +L_{\tau}$
and other linear
combination of these quantum numbers depending on the kind of neutrino
mixings that one wants to obtain. The general principle we will adopt is
that global symmetries will only be broken by soft
terms in the potential and any other 
interaction with dimension $d\geq 4$ will respect 
the symmetry.

As an application of our mechanism, we apply it to understand the lightness of
the sterile neutrino, that is required to provide a simultaneous fit to solar,
atmospheric and the LSND data~\cite{lsnd}.  
We find that in simple models that implement
our strategy,  the sterile neutrino, even though is a standard model gauge
singlet, may get a small mass comparable to that of familiar neutrinos due to
the higher dimensional operators, leading naturally, for instance,
to a (3+1) scenario which has recently been discussed~\cite{peres,bala}.
We must however mention that the (3+1) scheme is generically constrained such
that it may work only for certain domain of values of the mass difference
$\Delta m^2_{LSND}$~\cite{peres,bala}, and there are also  recent
papers~\cite{problems} noting that it may be lest favoured as the explanation
of all neutrino anomalies, though it has not been ruled out, and so, 
it remains of some interest.

\section{The new mechanism}

Before presenting our proposal, let us review the higher dimensional
mechanism suggested by Weinberg~\cite{weinberg}. According to this
mechanism, one writes down the nonrenormalizable operator involving the
standard model fields that can lead to neutrino masses,
\begin{eqnarray}
 {\cal L}= \frac{f_{ab}}{\Lambda}\left( L^T_{ia}C^{-1} L_{jb} H_k H_l 
\epsilon_{ik}\epsilon_{jl}\right),
\label{ge5do}
\end{eqnarray}
where $i,j,k,l$ are $SU(2)_L$ indices and $\Lambda$ is the scale of new
physics. This operator  breaks explicitly the lepton number. When $H$
develops a vacuum expectation value (vev), neutrinos acquire masses,
\begin{eqnarray}
m_\nu = f \frac{\langle H \rangle^2 }{\Lambda}.
\label{nmtefdo}
\end{eqnarray}
{}From Eq.~(\ref{nmtefdo}), it is clear that this formula is similar to the
seesaw formula in Eq. (1) with Dirac mass replaced by the Higgs vev that
determines the weak scale and to get neutrino masses in  eV range or
less implies the scale of new physics $\Lambda\geq 10^{13}$ GeV or so.
We should mention that a recent study~\cite{babu2} 
has classified several other non renormalizable 
operators of dimension
smaller than 11 that may give rise to neutrino masses. 
However those equivalent
to the ones  we will  use here were not considered there.

Let us now add to the SM representation content a singlet scalar field
$\chi$ which  carries lepton number $L=-1$ and a second scalar doublet
$H'$ with $L=0$. We then impose the discrete $Z_2$ symmetry on the model
under which $H'$ and $\chi$ are odd while all other fields are even. Then
the higher dimensional operator given above is forbidden by lepton number
conservation. The lowest dimensional operator that is invariant under $L$
is given by:
\begin{eqnarray}
{\cal L}=\frac{f}{M^3}(LH\chi)^2 +\frac{f'}{M^3}(LH'\chi)^2,
\label{eod6}
\end{eqnarray}
where $M$ is now the new scale of physics.
Once  $H,H'$ and $\chi$ develop vev's we obtain the following expression
to neutrino masses:
\begin{eqnarray}
 m_\nu = \frac{\langle \chi \rangle^2}{M^3}\left(
  f \langle H \rangle^2  + f' \langle H' \rangle^2 \right).
\label{nmeod6}
\end{eqnarray}
If the vev of $\langle\chi\rangle$ is suppressed, then $M$ can be in the TeV
range. We now show that this happens quite naturally in our model. 

To study the
vevs of the Higgs fields, we write down the most general Higgs potential
consistent with the symmetries we have imposed on the model, i.e., the gauge
symmetry and global $U(1)_L$. However to generate a natural small vacuum to 
$\chi$ we consider in the potential terms 
that break explicitly the lepton number,
\begin{eqnarray}
V(H,H',\chi)&=& {\mu^2_H} H^{\dagger}H +
\frac{\lambda_1}{2}(H^{\dagger}H)^2 
+{\mu^2_{H'}} H^{'\dagger}H' +
\frac{\lambda_2}{2}(H^{'\dagger}H')^2
+{\lambda_3}H^{\dagger}H H^{'\dagger}H' \nonumber \\
&& +\frac{\mu^2_\chi}{2} \chi^{\dagger} \chi +
\frac{\lambda'_2}{2}(\chi^{\dagger} \chi)^2 
+{\lambda_4}H^{\dagger}H\chi^{\dagger}\chi 
+{\lambda'_4}H^{'\dagger}H'\chi^{\dagger}\chi 
- MH^{\dagger}H'\chi + h.c.
\label{potential}
\end{eqnarray}
First note that the $M$ term in the potential breaks the global
$L$-symmetry softly.
For $\mu^2_{H,H'} < 0$, the potential leads to both $H$ and $H'$ fields
having nonzero vev's that break the electroweak symmetry. Minimization of the
potential gives for  the parameterization,
 \be 
 \left(\ba{c} \langle H\rangle\\ \langle H'\rangle \ea\right) = 
 v \left(\ba{c} \cos\beta\\ \sin\beta\ea\right),
 \ee
the mixing angle:
 \be 
 \tan^2\beta = {\lambda_2 \mu^2_H - \lambda_3 \mu^2_{H'}\over 
                 \lambda_1 \mu^2_{H'} - \lambda_3 \mu^2_H}.
 \ee
If one assumes
the $d_R$ and $e_R$ to be odd under the $Z_2$ symmetry, then one can
choose $\langle H'\rangle\ll \langle H\rangle$ 
as in the case of large $\tan\beta$ supersymmetric
models. For $\mu^2_{\chi} >0$ and $\lambda'_2 \geq 0$, the $M$ term in the
potential then induces a vev for the $\chi$ field given by:
\begin{eqnarray}
\langle\chi\rangle = \frac{M\langle H\rangle\langle H'\rangle}{M^2_\chi}\equiv
\frac{Mv_uv_d}{\mu^2_{\chi}+\lambda'_2\langle\chi^2\rangle + v^2
\cos\beta^2(\lambda_4 + \lambda'_4 \tan^2\beta)} .
\end{eqnarray}
If we assume $v_d\simeq \frac{v_u}{10}$ and $M_\chi\sim M\sim 20$ TeV, we get
$\langle\chi\rangle\simeq 0.3$ GeV, without any fine tuning of parameters. 
Note that
this leads to breaking of lepton number. There is however no zero mass
particle since the $M$ term in the potential breaks lepton number
explicitly. In fact the masses of the real fields in $\chi$ become of
order of the the scale $M$. It is worth noticing that this is nothing but a
singlet realization of the so called Type II see-saw mechanism~\cite{typeii}. 
Indeed in all our analysis above
one can use a electroweak triplet carrying the
same lepton number instead of the singlet $\chi$, with the richness on the
phenomenology of such a field. For what we want to discuss hereafter, whether
$\chi$ is a singlet or a triplet will not be relevant.

Using Eq.~(\ref{nmeod6}) and $f\sim 0.1$, we get $m_{\nu}\simeq 0.04$ eV. 
It is interesting that the muon neutrino mass required to explain the
atmospheric neutrino deficit emerges without any fine tuning of parameters.
This is the new mechanism which we want to exploit in the rest of the
paper to understand neutrino mass patterns. This mechanism has also the
potential to be useful in understanding small neutrino masses in low
string scale theories, where the highest available scale is in the multi
TeV range. Moreover, it may even generate a naturally 
small sterile neutrino mass. 
We should notice  that the present mechanism can also be 
seen as a Type II see-saw realization of the dimension 
nine operator $L^2H^6$. Notice that such class of operators 
was explicitly excluded  in the general analysis of Ref.~\cite{babu2}. 

Let us conclude this section describing an underlying theory that could lead to 
the higher dimensional operators as part of the low energy effective theory. 
One possibility is to have at the 10 TeV scale three new singlet fields
of mass of order 10 TeV: $N_{1,2,3}$; such that $N_{1,2}$ have $L=\pm 1$
respectively whereas $N_3$ has $L=0$ and is odd under $Z_2$. The invariant 
potential involving these fields is then given by:
\begin{eqnarray}
{\cal L}= LHN_2 + M_1N_1N_2 +N_3\chi N_1 + M_3 N_3N_3 + h.c.
\end{eqnarray}
$M_{1,3}$ are assumed to be of order 10 TeV. This theory leads in the low
energy limit to the effective theory described in~(\ref{eod6}) (see Fig. 1). 
We now move into considering 
how the  desired pattern for the neutrino masses can be obtained.

\section{Bimaximal mixing pattern}

In this section, we apply the mechanism of the previous section to derive
the bimaximal neutrino mixing pattern. For this purpose, we recall that in
the previous section, we used the softly broken global symmetry $L$.
Suppose, in addition, we have a softly broken global 
$L'= L_e-L_{\mu}-L_{\tau}$. 
Then, the allowed higher dimensional terms in the
Eq.~(\ref{nmeod6}) would have generation labels in them and would lead to a
mass matrix of the form:
\begin{eqnarray}
{\cal M}^{(0)}_{\nu}= m \left(\begin{array}{ccc} 0 & c_\theta & s_\theta\\
c_\theta  & 0 & 0 \\
s_\theta & 0 & 0\end{array}\right)~,
\label{m1}
\end{eqnarray}
where  $c_\theta$ ($s_\theta$) represents the function $\cos\theta$
($\sin\theta$).
It is well known that this leads to the bimaximal neutrino mixing
pattern (see for instance Refs.~\cite{gold,theory,jr,barb,more}):
\begin{eqnarray}
U_{BM}= \left(\begin{array}{ccc} \frac{1}{\sqrt{2}} & -\frac{1}{\sqrt{2}}
& 0 \\ \frac{c_\theta}{\sqrt{2}} & \frac{c_\theta}{\sqrt{2}} & -s_\theta 
\\ \frac{s_\theta}{\sqrt{2}}
& \frac{s_\theta}{\sqrt{2}} & c_\theta \end{array}\right). 
\end{eqnarray}
As it is evident from~(\ref{m1}), at this stage, there is no mass splitting
between $\nu_1$ and $\nu_{2}$ that can lead to oscillations. There is
however mass splitting between $\nu_{1,2}$ and $\nu_3$ with 
$\Delta m^2= m^2$, 
which give rise to oscillations with the amplitude $\sin^2 2\theta$.
As we saw before, for the parameters chosen in the
previous section, this is precisely in the range required to solve the
atmospheric neutrino anomaly, since then 
$\Delta m_{atm}^2=m^2\approx 1.6\times 10^{-3}~{\rm eV}^2$. 
The interesting point to reemphasize in this
connection is that the neutrino mass spectrum is an inverted one.

In order to generate the splitting between the $\nu_1$ and $\nu_2$, we
need to introduce soft breaking of $L' = L_e-L_{\mu}-L_{\tau}$ symmetry so
that an entry such as $m_{ee}$ or $m_{\mu\mu}$ can be generated. For solar
neutrino anomaly solution via the large angle MSW mechanism, one would
require $m_{ee}$ to be of order $10^{-4}$ eV and for resolution via the
vacuum oscillation solution, one would require $m_{ee}\sim 10^{-9}$ eV.
For this purpose it is convenient to introduce a field $\delta$ with $L=0$ but
$L'=-1$. The vev of this field can also be induced  by the mechanism developed 
in the last section, but this time we choose the parameters such that
$\langle\delta\rangle/M\sim 10^{-1}$. This allows us to introduce a higher
dimensional operator of the type:
\be
 (L_eH)^2\chi^2\delta^2/M^5.
 \label{split}
\ee
Once all vevs
are substituted, one gets $m_{ee}\sim 4\times 10^{-4}$ eV. This favors the
large angle MSW solution to the solar neutrino problem. In fact, the solar
mass parameter is then given as
$\Delta m_{\odot}^2 \approx 
2 m^2 (\langle\delta\rangle/M)^2\approx 3\cdot 10^{-5}~{\rm eV}^2$.
A smaller vacuum, 
$\langle\delta\rangle\sim 1$ GeV, however, will produce 
$m_{ee}\sim 3\times 10^{-9}$ eV for vacuum solution.

\section{The (3+1) scheme}

The above neutrino mass scheme does not explain the LSND results. In order
to include it into our model, the simplest possibility in our framework 
appears to be the so called (3+1) scheme, that may account for the explanation of
all neutrino anomalies~\cite{peres,bala,problems}. 
For this purpose, we include a left-handed singlet neutrino
in the model, $\nu_s$, that
carries its own lepton number $L_s$, and two
odd scalar singlets, besides $\chi$, which we call 
$\delta_1$ and $\delta_2$.
In order to get the final mass pattern, extra softly broken 
global symmetries are needed. Following our principle, we
take those symmetries to be linear combinations of the four lepton numbers: 
$L_{e,\mu,\tau,s}$. 
Let us consider, for instance,  three of such  
global symmetries: $U(1)_{L_1}$; $U(1)_{L_2}$ and $U(1)_{L_3}$, where 
$L_{1,2,3}$ represents our specific choice for the 
linear combinations of lepton numbers.  We have  taken
such symmetries among the total active lepton number, $L=L_e + L_\mu + L_\tau$;
the sterile lepton number, $L_s$; the total lepton number, $L_T= L + L_s$;
and the combinations:  $L'= L_e - L_\mu - L_\tau$ and $L''= L_s-L$.
We always pick up $L_2= L'$ to
maintain as much as possible our previous results. 
The quantum numbers associated to the chosen  symmetries 
and carried by the scalar fields are depicted in Table I, where three different
models are considered. 
All these models  have the same output, 
they provide the following higher
dimensional mass terms  (at the lower order)  that conserve such symmetries:

i) Active flavour  mass terms:
 \be 
 f_{ij} {L_i H L_j H \chi^2\over M^3} ,
 \ee
where the Yukawa couplings $f_{ij}$; for $i=e\mu\tau$,
are only constrained by the $L'$ symmetry. These
terms give rise to the same mass matrix as in Eq.~(\ref{m1}), and same order of
parameters, that now we assume. 

ii) Active-sterile couplings: 
 \be  
 g_e {L_e H' \nu_s \chi^2 \delta_1\over M^3}
 + \left(g_\mu L_\mu  + g_\tau L_\tau \right)
  {H' \nu_s  \chi^2\delta_2\over M^3}.
 \ee 
In this case the Yukawa couplings are all non zero, and if
$\langle\delta_{1,2}\rangle$ are both the same order then the mass couplings,
$\mu_i\sim \langle H\rangle\langle \chi\rangle^2\langle \delta\rangle/M^3$, 
are comparable among each other.  For simplicity one may take the light 
hierarchy  $g_e:g_\mu:g_\tau \approx 1:c_\theta:s_\theta$. 
This choice is suggested by the  fact that they are governed  by 
$L'$ as well as the active mass terms. Some amount of
tuning may be needed in our example, but small deviations of our choice will not
affect our conclusions.  However, this particular choice 
will allow us to make more explicit calculations later on.

iii) Finally, the sterile mass term:
 \be
  h {\nu_s \nu_s \chi^2 \delta_1 \delta_2\over M^3}.
 \ee 
That give rise to the sterile mass, 
$m_s\sim\langle\chi\rangle^2\langle\delta_1\rangle\langle\delta_2\rangle/M^3$. 
Now,  if we take  $\langle\chi\rangle\sim .3 $ GeV, as before with 
$M\sim 20$ TeV
and $\langle\delta\rangle\sim 400$ GeV, we get for $h\sim 1$,  
the sterile mass
$m_{s} \sim 2$ eV,  which is adequate to explain the mass 
difference required by  the LSND data. 

After introducing  expectation values we get, in the basis 
($\nu_\alpha,\nu_s$), the mass matrix,
 \be 
  {\cal M} = \left( \begin{array}{c c}
           {\cal M}^{(0)} &  {\vec\epsilon}~m_{s}\\
	{\vec\epsilon}\,^\dagger~ m_{s} & m_{s}
       \end{array} \right) ~,
  \label{m31}
  \ee
where 
$\epsilon_i = \mu_i/m_s \sim (g_i/h)\langle H\rangle/\langle\delta\rangle$. 
Thus, by taking $g\sim {\cal O}(1)$, this give us the hierarchy on masses,
$m\ll \mu_i < m_s$ that will reproduce the features of the  (3+1) scheme.

Let us now analyze how the required parameters appear from Eq.~(\ref{m31}).
First thing to notice is that at the lower order in $\epsilon$, the unitary 
mixing matrix has the form:
 \be 
U = 
\left(\ba{cc} U_{BM} & \vec\epsilon \\ 
{\vec\epsilon}\,^\dagger U_{BM} & 1 \ea\right).
 \label{umix0}
 \ee
Therefore, as an immediate conclusion we have that $U_{e3}=0$, as desired by
CHOOZ. Also, LSND arises with 
$\sin^2\theta_{LSND}= 4 |\epsilon_e \epsilon_\mu|^2$, that
constrains the active to sterile couplings to values well 
in our preferred range of parameters ($O(10^{-1})$).

Next, while the sterile mass remains almost unperturbed, 
all other mass terms in the active block  get a see-saw type
correction~\cite{bala},
 \be 
{\cal M}^{(0)} \rightarrow {\cal M}^{(0)} - 
{\vec\epsilon}~{\vec\epsilon}~^\dagger m_s~.
\label{}
\ee
The effect of this correction is to break the degeneracy of  $\nu_1$ and
$\nu_2$ induced by $L'$, 
since in general it  introduces small diagonal terms (among
others). Thus, a nice feature of this model is that  we do not need to invoke
any other mechanism to generate those corrections. They come naturally due to
the breaking of the other symmetries.

Let us now consider as an specific 
example the above suggested hierarchy of the $g$
Yukawa couplings. We then take 
${\vec\epsilon}~^\dagger \sim 
(|{\vec\epsilon}|/\sqrt{2}) (1,c_\theta,s_\theta)$.
After rotating the active sector by $U_{BM}$, ${\cal M}^{(0)}$ 
becomes $Diag(m,-m,0)$,
where $m=\sqrt{\Delta m_{atm}^2}\approx 0.04$ eV, as before. 
The same rotation projects ${\vec\epsilon}$ along the direction
$(1,0,0)$.
Hence, $\nu_3$ remain exactly  massless, 
and the 1-4 sector remains mixed with a mixing angle
$\tan\beta \approx |{\vec\epsilon}|$. Therefore, after a complete 
diagonalization,  one gets the exact mixing matrix:
 \be
 U_{mix}= \left(\begin{array}{cccc} {c_\beta\over\sqrt{2}} & {-1\over\sqrt{2}}
 & 0  & {s_\beta\over\sqrt{2}}\\ 
 \frac{c_\theta c_\beta}{\sqrt{2}} & \frac{c_\theta}{\sqrt{2}} & 
 -s_\theta & {c_\theta s_\beta\over \sqrt{2}} \\ 
 \frac{s_\theta}{\sqrt{2}} & \frac{s_\theta}{\sqrt{2}} & 
 c_\theta & {s_\theta s_\beta\over \sqrt{2}} \\
 -s_\beta & 0 & 0 & c_\beta 
 \end{array}\right).
 \label{umix}
 \ee
The mass spectrum we then get is, 
 \bea
 m_1 &=& m - |{\vec\epsilon}|^2 m_s ~, \nonumber\\
 m_2 &=& -m ~,\nonumber\\
 m_3 &=& 0 ~,\nonumber\\
 m_4 &=& m_s(1+|{\vec\epsilon}|^2) ~.
 \label{masses}
 \eea

Let us see how the solar neutrino mass arise. Looking at the spectrum
(\ref{masses}), we require
 $\Delta m_{21}^2 = m_2^2 - m_1^2 = \Delta m^2_{\odot}$,
that renders,
\be
 |{\vec\epsilon}|^2 \approx {2m\over m_s} - {\Delta m^2_{\odot} \over 2 m m_s}~.
\ee
This may imply a quite large cancellation on the parameters of the theory, 
and is a manifestation of the constrained solution we are discussing. The
required fine tunning would, however, be not too large. 
For large mixing angle solutions one needs that $|\vec\epsilon|^2$ be adjusted
at the level of 10\% at most, such that the difference
$2m -|\vec\epsilon|^2$ be of the order of  $10^{-3}$.

By taking $m_s\sim 1-2$ eV as the LSND scale 
(just as it came from our previous general analysis), 
for the large mixing MSW 
solution of solar neutrino problem, one gets: 
$|{\vec\epsilon}|^2 \approx 0.04 - 0.08$. 
Therefore, our prediction for the LSND oscillation probability is
\be 
\sin^22\theta_{LSND}= \cos_\theta^2 |{\vec\epsilon}|^4
\approx (.8 - 3.2) \times 10^{-3} ,
\ee
where the extreme right hand side has been evaluated assuming maximal mixing in
the active sector. We should comment that, as already observed in the
literature~\cite{peres,bala}, this also pushes Bugey and CDHS to the limit. 
With
our set of parameters we get: $\sin^22\theta_{\rm Bugey}\sim 0.04-0.08$ and 
$\sin^22\theta_{\rm CDHS}\sim 0.08-0.16$. 
On the other hand, this could also be
interpreted as a chance to have  accessible signals of $\nu_{e,\mu}$
disappearance in future experiments.  
Given our (inverted) hierarchy,  other mixing parameters are  given as
\bea 
\sin^22\theta_{atm}&=& 
\sin^22\theta\cdot \left(1 - {1\over 2}|{\vec\epsilon}|^2 \right)
\approx 0.96 - 0.98~, \nonumber \\
\sin^22\theta_{\odot}&=& (1- |{\vec\epsilon}|^2)\approx 0.92 - 0.96~,
\eea
which means that solar and atmospheric are both explained by large mixing angle
solutions in our model.

\section{concluding remarks}

We have presented a new mechanism that may  generate  small neutrino masses at
tree level which does not need the existence of  large scales for new physics
but rather make  use of higher dimensional operators generated at the TeV 
scale. In the realization of our mechanism the smallness of the neutrino masses
are due to the large mass suppression on  dimension seven  operators which are
involved in the generation of masses and due to the presence of a relatively
small vacuum of a scalar particle which (softly) breaks down lepton number. 
Such a  vacuum arises from a singlet scalar realization of the so called Type II
see-saw mechanism. As a direct application we have explored the generation of
textures that  may explain the neutrino anomalies. Given the flavour symmetry 
$L^{\prime}: L_e -L_\mu -L_\tau$ we generate the bimaximal neutrino mixing
patter, which we further extended by adding extra symmetries and 
a sterile neutrino which also 
obtains a eV mass, giving rise to the (3+1) scenario that may account for
including LSND in the scheme. Our mechanism may be useful in theories where the
fundamental (string) scale is in the TeV range, since it relays only on
brane physics for the generation of neutrino masses.

\vskip2em

{\it Acknowledgements.} 
We would like to thank R.N. Mohapatra for several  valuable comments and
stimulating discussions. The work of CP is supported by Funda\c c\~ao de
Amparo \`a Pesquisa do Estado de S\~ao Paulo (FAPESP).


\begin{table}
\begin{tabular}{||c| c | c| c|| c| c| c || c| c| c || }
&\multicolumn{3}{c||}{Model I} & \multicolumn{3}{c||}{Model II} & 
 \multicolumn{3}{c ||}{Model III} \\
 \hline
  Field    & $L_T$ & $L'$ & $L''$& $L_T$ & $L'$ & $L_s$&
   $L$ & $L'$ & $L_s$\\
\hline
$\chi $    & -1  &  0  &  1 &-1 & 0 & 0&-1 & 0 & 0\\
$\delta_1$ &  0  & -1  & -2 & 0 &-1 &-1& 1 &-1 &-1\\
$\delta_2$ &  0  &  1  & -2 & 0 & 1 &-1& 1 &-1 & 1
\end{tabular}
\vskip1ex

\caption{Assignments of charges for the scalar fields in 
three different models with the  flavour symmetries as shown, which 
give rise to the (3+1) scenario. 
Here $L_T = L_s + L$; $L=L_e + L_\mu + L_\tau$; 
$L'=L_e - L_\mu - L_\tau$; and $L'' = L_s - L$.
Standard Higgs doublets are in all cases chargeless under this symmetries.}
\end{table}

\begin{figure}
\centerline{
\epsfxsize=300pt
\epsfbox{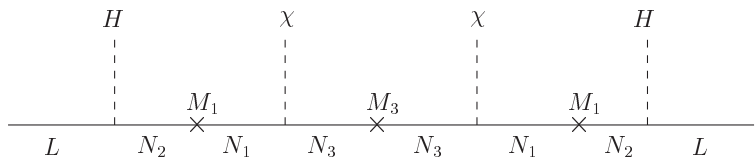}
}
\vskip1ex

\caption{Tree-level realization of the effective operator in Eq. (\ref{eod6}).}
\end{figure}

\end{document}